# A Robust Logical and Computational Characterisation of Peer-to-Peer Database Systems

Enrico Franconi[1], Gabriel Kuper[2], Andrei Lopatenko[13], and Luciano Serafini[4]

[1] Free University of Bozen–Bolzano, Faculty of Computer Science, Italy,
franconi@inf.unibz.it, alopatenko@unibz.it
[2] University of Trento, DIT, Italy, kuper@acm.org
[3] University of Manchester, Department of Computer Science, UK
[4] ITC-irst Trento, Italy, serafini@itc.it

**Abstract.** In this paper we give a robust logical and computational characterisation of peer-to-peer (p2p) database systems. We first define a precise model-theoretic semantics of a p2p system, which allows for local inconsistency handling. We then characterise the general computational properties for the problem of answering queries to such a p2p system. Finally, we devise tight complexity bounds and distributed procedures for the problem of answering queries in few relevant special cases.

## 1 Introduction

The first question we have to answer when working on a logical characterisation of p2p database systems is the following: what is a p2p database system in the logical sense? In general, it is possible to say that a p2p database system is an integration system, composed by a set of (distributed) databases interconnected by means of some sort of logically interpreted mappings. However, we also want to distinguish p2p systems from standard classical logic-based integration systems, as for example described in [?]. As a matter of fact, a p2p database system should be understood as a collection of independent nodes where the *directed* mappings between nodes have the only role to define how data migrates from a set of source nodes to a target node. This idea has been already clearly formulated in [?], where a framework based on KFOL is informally proposed as a possible solution.

Consider the following example. Suppose we have three distributed databases. The first one ($DB_1$) is the municipality's internal database, which has a table `Citizen-1`. The second one ($DB_2$) is a public database, obtained from the municipality's database, with two tables `Male-2` and `Female-2`. The third database ($DB_3$) is the Pension Agency database, obtained from a public database, with the table `Citizen-3`. The three databases are interconnected by means of the following rules:

$1 : \texttt{Citizen-1}(x) \Rightarrow 2 : (\texttt{Male-2}(x) \vee \texttt{Female-2}(x))$
  (this rule connects $DB_1$ with $DB_2$)

$\quad 2 : \texttt{Male-2}(x) \Rightarrow 3 : \texttt{Citizen-3}(x)$
$\quad 2 : \texttt{Female-2}(x) \Rightarrow 3 : \texttt{Citizen-3}(x)$
$\quad\quad$ (these rules connect $DB_2$ with $DB_3$)

In the classical logical model, the `Citizen-3` table in $DB_3$ should be filled with all of the individuals in the `Citizen-1` table in $DB_1$, since the following rule is logically implied:

$\quad 1 : \texttt{Citizen-1}(x) \Rightarrow 3 : \texttt{Citizen-3}(x)$

However, in a p2p system this is not a desirable conclusion. In fact, rules should be interpreted only for fetching data, and not for logical computation. In this example, the tables `Female-2` and `Male-2` in $DB_2$ will be empty, since the data is fetched from $DB_1$, where the gender of any specific entry in `Citizen-1` is not known. From the perspective of $DB_2$, the only thing that is known is that each citizen is in the view (`Female-2` ∨ `Male-2`). Therefore, when $DB_3$ asks for data from $DB_2$, the result will be empty.

In other words, the rules

$\quad 2 : \texttt{Male-2}(x) \Rightarrow 3 : \texttt{Citizen-3}(x)$
$\quad 2 : \texttt{Female-2}(x) \Rightarrow 3 : \texttt{Citizen-3}(x)$

will transfer no data from $DB_2$ to $DB_3$, since no individual is known in $DB_2$ to be either definitely a male (in which case the first rule would apply) or definitely a female (in which case the second rule would apply). We only know that any citizen in $DB_1$ is either male or female in $DB_2$, and no reasoning about the rules should be allowed.

$\quad$ We shall give a robust logical and computational characterisation of p2p database systems, based on the principle sketched above. We say that our formalisation is *robust* since, unlike other formalisations, it allows for local inconsistencies in some node of the p2p network: if some database is inconsistent it will not result in the entire database being inconsistent. Furthermore, we propose a polynomial-time algorithm for query answering over realistic p2p networks, which does not have to be aware of the network structure, which can therefore change dynamically.

$\quad$ Our work has been influenced by the semantic definitions of [?], which itself is based on the work of [?]. [?] defined the *Local Relational Model* (LRM) to formalise p2p systems. In LRM all nodes are assumed to be relational databases and the interaction between them is described by coordination rules and translation rules between data items. Coordination rules may have an arbitrary form and allow to express constraints between nodes. The model-theoretic semantics of coordination rules in [?,?] is non-classical, and it is very close to the *local semantics* introduced in this paper.

$\quad$ Various other problems of data management focusing on p2p systems have been considered in the literature with classical logic-based solutions. We mention here only few of them. In [?], query answering for relational database- based p2p systems under classical semantics is considered. The case when both GAV and LAV style mappings between peers are allowed is considered. The mapping between data sources is given in the $\mathcal{PPL}$ language allowing for both inclusion and equality of conjunctive queries over data sources and definitional mappings (that is, inclusions of positive queries for a relation), and queries have certain answer semantics. It is proved that in the general case query answering is undecidable and in the acyclic case with only inclusion mappings allowed, the complexity of query answering becomes polynomial (if equality peer mappings are allowed, subject to some restrictions, query answering then becomes co-NP-complete). An algorithm reformulating a query to a given node into queries to nodes containing data is provided. In [?] mapping tables (similar to translation rules of [?]) are considered. In the article mapping tables under different semantic are considered, as well as constraints on mappings and reasoning over tables and constraints under such conditions. Moreover, see [?] for the data placement problem, [?] for data trading in data replication, [?] for the relationship between p2p and Semantic Web, and in general [?] for the best survey of classical logic-based data integration systems.

This paper is organised as follows. At the beginning, the formal framework is introduced; three equivalent ways of defining the semantics of a p2p system will be given, together with a fourth one – the extended local semantics – which is able to handle inconsistency and will be adopted in the rest of the paper. General computational properties will be analysed in Section 3, together with the special case of p2p systems with the minimal model property. Tight data and node complexity bounds for query answering are devised for the Datalog-p2p systems and for the acyclic p2p systems.

## 2  The Basic Framework

We first define the nodes of our p2p network as general first order logic (FOL) theories sharing a common set of constants. Thus, a node can be seen as represented by the set of models of the FOL theory.

**Definition 1 (Local database)** *Let $I$ be a nonempty finite set of indexes $\{1, 2, \ldots, n\}$, and $C$ be a set of constants. For each pair of distinct $i$, $j \in I$, let $L_i$ be a first order function-free language with signature disjoint from $L_j$ but for the shared constants $C$. A* local database $DB_i$ *is a theory on the first order language $L_i$.*

Nodes are interconnected by means of coordination rules. A coordination rule allows a node $i$ to fetch data from its neighbour nodes $j_1, \ldots, j_m$.

**Definition 2 (Coordination rule)** *A* coordination rule *is an expression of the form*
$$j_1 : b_1(\mathbf{x}_1, \mathbf{y}_1) \wedge \cdots \wedge j_k : b_k(\mathbf{x}_k, \mathbf{y}_k) \Rightarrow i : h(\mathbf{x})$$
*$j_1, \ldots, j_k, i$ are distinct indices, and each $b_l(\mathbf{x}_l, \mathbf{y}_l)$ is a formula of $L_{j_l}$, and $h(\mathbf{x})$ is a formula of $L_i$, and $\mathbf{x} = \mathbf{x}_1 \cup \cdots \cup \mathbf{x}_k$.*

Please note that we are making the simplifying assumption that the equal constants mentioned in the various nodes are actually referring to equal objects, i.e., they are playing the role of URIs (Uniform Resource Identifiers). Other approaches consider *domain relations* to map objects between different nodes [**?**]. We will consider this extension in our future work.

A p2p system is just the collection of nodes interconnected by the rules.

**Definition 3 (p2p system)** *A* peer-to-peer (p2p) system *is a tuple of the form* $MDB = \langle LDB, CR \rangle$, *where* $LDB = \{DB_1, \cdots, DB_n\}$ *is the set of local databases, and CR is the set of coordination rules.*

A user accesses the information hold by a p2p system by formulating a query to a specific node.

**Definition 4 (Query)** *A* local query *is a first order formula in the language of one of the databases* $DB_i$.

## 2.1 Global Semantics

In this section we formally introduce the meaning of a p2p system. We say that a global model of a p2p system is a FOL interpretation over the union of the FOL languages satisfying both the FOL theories local to each node and the coordination rules. Here it is crucial the fact that the semantics of the coordination rule is not the expected standard universal material implication, as in the classical information integration approaches. The p2p semantics for the coordination rules states that if the body of a rule is true in any possible model of the source nodes then the head of the rule is true in any possible model of the target node. This different notion from classical first order logic is exactly what we need: in fact, only information which is true in the source node is propagated forward.

**Definition 5 (Global semantics)** *Let $\Delta$ be a non empty set of objects including $C$ (see Definition 1), and let $MDB = \langle LDB, CR \rangle$ be a p2p system. An* interpretation *of MDB over $\Delta$ is a n-tuple $m \equiv \langle m_1, m_2, \ldots m_n \rangle$ where each $m_i$ is a classical first order logic interpretation of $L_i$ on the domain $\Delta$ that interprets constants as themselves.*
*We adopt the convention that, if $m$ is an interpretation, then $m_i$ denotes the $i^{th}$ element of $m$.*
*A* (global) model *$M$ for $MDB$ – written $M \models_{\text{global}} MDB$ – is a nonempty set of interpretations such that:*

1. *the model locally satisfies the conditions of each database, i.e.,*

$$\forall m \in M.\ (m_i \models DB_i)$$

2. *and the model satisfies the coordination rules as well, i.e., for any coordination rule*
$$j_1 : b_1(\mathbf{x}_1, \mathbf{y}_1) \wedge \cdots \wedge j_k : b_k(\mathbf{x}_k, \mathbf{y}_k) \Rightarrow i : h(\mathbf{x})$$

*then for every assignment $\alpha$ – assigning the variables $\mathbf{x}$ to elements in $\Delta$, which is common to all models – the following holds:*

$$(\forall m \in M.(m_{j_1} \models \exists \mathbf{y}.b_1(\mathbf{x}_1, \mathbf{y})) \wedge \cdots \wedge (m_{j_k} \models \exists \mathbf{y}.b_k(\mathbf{x}_k, \mathbf{y}))) \rightarrow$$
$$(\forall m \in M.\ (m_i \models h(\mathbf{x})))$$

The answer to a query in a node of the system is nothing else than the tuples of values that, substituted to the variables of the query, make the query true in each global model restricted to the node itself.

**Definition 6 (Query answer)** *Let $Q_i(\mathbf{x})$ be a local query with free variables $\mathbf{x}$. The* answer set *of $Q_i$ is the set of substitutions of $\mathbf{x}$ with constants $\mathbf{c}$, such that any model $M$ of MDB satisfies the query, i.e.,*

$$\{\mathbf{c} \in C \times \cdots \times C \mid \forall M.\ (M \models_{\text{global}} MDB) \rightarrow \forall m \in M.\ (m_i \models Q_i(\mathbf{c}))\}$$

*This corresponds to the definition of certain answer in the information integration literature.*

### 2.2 Local Semantics

The semantics we have introduced in the previous section is called global since it introduces the notion of a global model which spans over the languages of all the nodes. In this section we introduce the notion of local semantics, where actually models of a p2p system have a node-centric nature which better reflects the required characteristics. We will prove at the end of the Section that the two semantics are equivalent.

**Definition 7** *The* derived local model $\hat{M}_i$ *is the union of the $i^{th}$ components of all the models of MDB:*

$$\hat{M}_i = \bigcup_{\substack{m \in M, \\ M \models_{\text{global}} MDB}} m_i$$

**Lemma 1** *The answer set of a local query $Q_i(\mathbf{x})$ coincides with the following:*

$$\{\mathbf{c} \in C \times \cdots \times C \mid \forall m_i \in \hat{M}_i.\ (m_i \models Q_i(\mathbf{c}))\}$$

The above lemma suggests that we could consider somehow $\left\langle \hat{M}_1, \ldots, \hat{M}_n \right\rangle$ as a model for the p2p system. This alternative semantics, which we call local semantics as opposed to the global semantics defined in the previous section, is defined in the following. The notation will sometimes coincide with the one used in the definition of global semantics; its meaning will be clear from the context.

**Definition 8 (Local semantics)** *A* (local) model $M$ for MDB – written $M \models MDB$ – is a sequence $\langle M_1, \ldots, M_n \rangle$ such that:

1. each $M_i$ is a non empty set of interpretations of $L_i$ over $\Delta$
2. $\forall m_i \in M_i.\ (m_i \models DB_i)$
3. for any coordination rule

$$j_1 : b_1(\mathbf{x}_1, \mathbf{y}_1) \wedge \cdots \wedge j_k : b_k(\mathbf{x}_k, \mathbf{y}_k) \Rightarrow i : h(\mathbf{x})$$

then for each assignment $\alpha$ to the variables $\mathbf{x}$ the following holds:

$$(\forall m_{j_1} \in M_{j_1}.(m_{j_1} \models \exists \mathbf{y}.b_1(\mathbf{x}_1, \mathbf{y}))) \wedge \cdots \wedge$$
$$(\forall m_{j_k} \in M_{j_k}.(m_{j_k} \models \exists \mathbf{y}.b_k(\mathbf{x}_k, \mathbf{y})))) \rightarrow$$
$$(\forall m_i \in M_i.\ (m_i \models h(\mathbf{x}))$$

**Definition 9 (Query answer for local semantics)** *Let $Q_i$ be a local query. The answer for $Q_i$ is the set of substitutions of $\mathbf{x}$ with constants $\mathbf{c}$ such that any model $M$ of $MDB$ locally satisfies the query, i.e.:*

$$\{\mathbf{c} \in C \times \cdots \times C \mid \forall M.\ (M \models MDB) \rightarrow \forall m_i \in M_i.\ (m_i \models Q_i(\mathbf{c}))\}$$

**Theorem 2** *The answer sets of a local query $Q_i$ in the global semantics and in the local semantics coincide.*

A way to understand the difference between global and local semantics would be the following. If

$$M = \{\langle m_1^1, \ldots, m_i^1, \ldots, m_n^1 \rangle, \ldots, \langle m_1^j, \ldots, m_i^j, \ldots, m_n^j \rangle, \ldots\}$$

is a model for a p2p system in the global semantics, then *also*

$$M' = \{\langle m_1^1, \ldots, m_i^j, \ldots, m_n^1 \rangle, \ldots, \langle m_1^j, \ldots, m_i^1, \ldots, m_n^j \rangle, \ldots\}$$

is a model in the global semantics. In other words, there is no formula expressible in the p2p system which distinguishes two models in the global semantics obtained by swapping local models. This is the reason why we can move to the local semantics defined in this section without loss of meaning. In fact, the local semantics itself does not distinguish between the two above cases, and can be therefore considered closer to the intended meaning of the p2p system.

### 2.3 Autoepistemic Semantics

In this section we briefly introduce a third approach to define the semantics of a p2p system, as suggested in [?]. This approach can be proved equivalent to the global semantics introduced at the beginning – and therefore equivalent to the local semantics as well.

Let us consider KFOL, i.e., the autoepistemic extension of FOL (see, e.g., [?]). The previous definition of global semantics can be easily changed to fit in a KFOL framework, so that the p2p system would be expressed in a single KFOL theory $\Sigma$. Each $D_i$ would be expressed into KFOL without any change, i.e., without using at all the **K** operator; the coordination rules would be translated into formulas in $\Sigma$ as

$$\forall \mathbf{x}.\mathbf{K}\exists \mathbf{y}.b(\mathbf{x},\mathbf{y}) \Rightarrow \mathbf{K}h(\mathbf{x}).$$

It can be easily proved that the answer set as defined above (Definition 6) in the global semantics framework is equivalent to the answer set defined in KFOL as the set of all constants **c** such that

$$\Sigma \models_K \mathbf{K}Q_i(\mathbf{c}) .$$

### 2.4 Extended Local Semantics to Handle Inconsistency

The semantics defined above does not formalise local inconsistency. In fact as soon as a local database becomes inconsistent, or a coordination rule pushes inconsistency somewhere, both the global and the local semantics say that no model of *MDB* exists. This means that local inconsistency implies global inconsistency, and the p2p system is not robust.

**Proposition 3** *For any p2p system such that there is an $i$ such that $DB_i$ is inconsistent, then the answer set of any query $Q_j(\mathbf{x})$ is equal to $C \times \cdots \times C$, for both the global and local semantics.*

In order to have a robust p2p system able to be meaningful even in presence of some inconsistent node, we extend the local semantics by allowing single $M_i$ to be the empty set. This captures the inconsistency of a local database: we say that a local database $DB_i$ is inconsistent if $M_i$ is empty for any model of the p2p system. A database depending on an inconsistent one through some coordination rule will have each dependent view – i.e., the formula in the head of the rules with $n$ free variables – equivalent to $\Delta^n$, and the databases not depending on the inconsistent one will remain consistent. Therefore, in presence of local inconsistency the global p2p system remains consistent.

The following example will clarify the difference between the local semantics and the extended local semantics in handling inconsistency.

*Example 1.* Consider the p2p system composed of a node $DB_1$ containing a unary predicate $P$ and an inconsistent axiom $\bot$, and another node $DB_2$ containing two unary predicates $Q$ and $R$ with no specific axiom on them. Let

$$1 : P(x) \Rightarrow 2 : Q(x)$$

be a coordination rule from $DB_1$ to $DB_2$. Even though $DB_1$ is inconsistent, there is a model $M = \langle M_1, M_2 \rangle$ where $M_2$ is not the empty set. The answer set of the query $Q(x)$ in 2 is the whole set of constants known to the p2p system. Furthermore, the answer set of the query $R(x)$ in 2 is the empty set. So, in this case the inconsistency does not have an effect through the coordination rule to each predicate of $DB_2$.

Let us suppose now that $M_2$ contains in addition the axiom $\exists x \neg Q(x)$. Then, the only model (in the local semantics) is $\langle M_1, M_2 \rangle$ where both $M_1$ and $M_2$ are the empty set.

In the case of fully consistent p2p systems, the local semantics and the extended local semantics coincide. In the case of some local inconsistency, the local (or, equivalently, the global) semantics will imply a globally inconsistent system, while the extended local semantics is able to still give meaningful answers.

**Theorem 4** *If there is a model for MDB with the local (or global, or autoepistemic) semantics then for each query the answer set with the local (or global, or autoepistemic) semantics coincide with the answer set with extended local semantics.*

## 3   Computing Answers

In this section, we will consider the global properties of a generic p2p system: we will try to find the conditions under which a computable solution to the query answering problem exists, we will investigate its properties and how to compute it in some logical database language. From now on, we assume the extended local semantics – i.e., the semantics of the p2p system able to cope with inconsistency. We include the sketches of some proofs.

Let us define the inclusion relation between models of a p2p system. A model $M$ is *included* into $N$ ($M \subseteq N$) if for each node $i$, a set of models of $i$ in $M$ is a subset of a set of models for $i$ in $N$.

Let $CR$ be a set of coordination rules and $M$ an interpretation of $MDB$, i.e., a sequence $\langle M_1, \ldots, M_n \rangle$ such that each $M_i$ is a set of interpretations of $L_i$ over $\Delta$. A ground formula $A$ is a *derived fact* for $M$ and $CR$ if either $M \models A$, or $i : \psi \Rightarrow j : A$ is an instantiation of a rule in $CR$ and $M \models \psi$. Please remember that when we write $M \models \psi$ – where $M$ is a model for $MDB$– we intend the logical implication for the extended local semantics.

**Definition 10 (Immediate consequence operator)** *Let MDB be a p2p system, CR a set of coordination rules, and M a model of MDB. A model $\hat{M}$ is an immediate consequence for M and CR if it is a maximal model included into M such that each $M_i \in \hat{M}$ contains facts derived by CR from M. The immediate consequence operator for MDB, denoted $T_{MDB}$, is the mapping from a set of models into a set of models such that for each $M$, $T_{MDB}(M)$ is an immediate consequence of $M$.*

Few lemmas about the properties of the consequence operator are in order to prove our main theorem.

**Lemma 5** *The operator $T_{MDB}$ is monotonic with respect to model inclusion, i.e., if $M \subseteq N$, then $T_{MBD}(M) \subseteq T_{MDB}(N)$*

*Proof.* For each rule create a ground instantiation of it. Each ground instance of $CR$ in $N$ is also present in $M$. This means that for each new formula $\psi$ derivable in $N$ the same formula is derivable in $M$. So, all models which are refused during the application of the operator in $N$ are also refused in $M$. Therefore, $T_{MDB}(M) \subseteq T_{MDB}(N)$.

**Lemma 6** *The operator $T_{MDB}$ is monotonic with respect to the set of ground instantiations of rules satisfied (the set of ground instances of rules derived at some step of the execution of an operator remains valid for all the subsequent steps).*

*Proof.* Let's assume that a rule $i : \psi(\mathbf{x},\mathbf{y}) \Rightarrow j : \phi(\mathbf{x})$ is instantiated for some $\mathbf{x}$, $\mathbf{y}$ at step $n$ for the set of models $M_i^n, M_j^n$. Clearly, it will remain valid for any step $m > n$, given the semantics of the rules and that $M_i^m \subseteq M_i^n, M_j^m \subseteq M_j^n$.

**Lemma 7** *For any initial model $M$, the operator $T_{MDB}$ reaches a fixpoint which is a model of MDB.*

*Proof.* Since we begin from a finite set of models, after a finite number of steps we reach a lower bound (possibly the empty set of models): this is a set of models which satisfy *MDB*. In fact, all local FOL theories are satisfied by definition of $T_{MDB}$, and if some rule in *CR* is not satisfied then an execution of $T_{MDB}$ will lead to a new model, but this would contradict the reaching of the fixpoint. If the empty set of models is reached then *MDB* is trivially satisfied.

The main theorem states that we can use the consequence operator to compute the answer to a query to a p2p system.

**Theorem 8** *The certain answer of a query to a p2p system MDB is the certain answer of the query over the model $T_{MDB}^\omega(M_0)$, where $M_0$ is the model set consisting of the Cartesian product of all the interpretations satisfying the local FOL theories.*

*Proof.* $\Leftarrow$. If $Q(a)$ is a certain answer, then, since $Q(a)$ is true in any model, it is true in the model resulting by applying the operator to the maximum original set. So, $\{\mathbf{x} \mid MDB \models Q_(\mathbf{x})\} \subseteq \{\mathbf{x} \mid T_{MDB}(M_0) \models Q_\mathbf{x}\}$

$\Rightarrow$. Since the original interpretation is the Cartesian product of all local interpretations, then any particular model consisting of a set of local models is a subset of $M_0$, i.e., $\forall M.M \subseteq M_0$. By monotonicity of the operator, it holds that

$$\forall M.T_{MDB}^\omega(M) \subseteq T_{MDB}^\omega(M_0)$$

Therefore, $\{\mathbf{x} \mid MDB \models Q(\mathbf{x})\} \supseteq \{\mathbf{x} \mid T_{MDB}(M_0) \models Q(\mathbf{x})\}$.

## 3.1 Computation with Minimal Models

Let us now assume that at each node the minimal model property holds – i.e., in each local database the intersection of all local models is a model itself of the local FOL theory, and it is minimal wrt set inclusion. Let us assume also that the coordination rules are preserving this property – e.g., the body of any rule is a conjunctive query and the head of any rule is a conjunctive query without existential variables. We say that in this case the p2p system enjoys the minimal model property. Then, it is possible to simplify the computation procedure defined by the $T_{MDB}$ operator. In such case the computation is reducible to a "migration of facts". The procedure is crucially simplified if it is impossible to get inconsistency in local nodes (like for Datalog or relational databases).

**Definition 11 (Minimal model property)** *The consequence operator $T_{MDB}^{min}$ for MDB with the minimal model property is defined in the following way:*

- *at the beginning, the minimal model is given for each node;*
- *at each step, $T_{MDB}^{min}$ computes for each coordination rule a set of derived facts and adds them into the local nodes;*
- *if for a node j an inconsistent theory is derived, then the current model is replaced by the empty set, otherwise the current theory is extended with the derived facts and the minimal model is replaced by the minimal model of the new theory.*

We denote with $T_{MDB}^{min,\omega}$ the fixpoint of $T_{MDB}^{min}$.

**Theorem 9** *If the p2p system has the minimal model property, then for positive queries Q(**x**)*

$$T_{MDB}^{min,\omega}(M_{min}) \models Q(\mathbf{x}) \quad \leftrightarrow \quad MDB \models Q(\mathbf{x})$$

*Proof.* If $M_{min}$ is the minimal model, then if $\psi$ does not contain negation, $(\forall M \text{ model of } MDB, M \models \psi) \Leftrightarrow M_{min} \models \psi$. Let us assume that we execute $T_{MDB}(M_0)$, where $M_0$ is the set of all the models of each node. Assume that at step $i$ of the execution of $T_{MDB}^{min}(M_{min})$ we get the minimal model of the outcome of step $i$ of the execution of $T_{MDB}(M_0)$ (which is evidently true for step 0). The set of derived facts for each node at step $i+1$ for $T_{MDB}$ will be the same as for $T_{MDB}^{min}$, so that at step $i+1$ the theories for the execution of $T_{MDB}$ and $T_{MDB}^{min}$ will be the same. By definition of $T_{MDB}^{min}$, this will give a minimal model at the $i+1$ step. If at step $n$ $T_{MDB}$ reaches a fixpoint, then $T_{MDB}^{min}$ reaches a fixpoint as well with the minimal model corresponding to the models devised by $T_{MDB}$. Since $Q$ is a positive query, the thesis is proved.

This theorem means that a p2p system with nodes and coordination rules with the minimal model property collapses to a traditional p2p and data integration system like [?,?] based on classical logic. A special case is when each node is either a pure relational database or a Datalog-based deductive database (in either case the node enjoys the minimal model property), and each rule has the body in the form of a conjunctive query and the head in the form of a conjunctive query without existential variables. We call such a system a Datalog-p2p system. In such case, it is possible to introduce a simple "global program" to answer queries to the p2p system. The global program is a single Datalog program obtained by taking the union of all local Datalog programs and of the coordination rules expressed in Datalog, plus the data at the nodes seen as EDB.

We are able to precisely characterise the data and node complexity of query answering in a Datalog-p2p system. The data complexity is the complexity of evaluating a fixed query in a p2p system with a fixed number of nodes and coordination rules over databases of variable size – as input we consider here the total size of all the databases. The node complexity, which we believe is a relevant complexity measure for a p2p system, is the complexity of evaluating a fixed query over a databases of a fixed size with respect to a variable number of nodes in a p2p system with a fixed number of coordination rules between each pair of nodes. It turns out that the worst case node complexity is rather high.

**Theorem 10 (Complexity of Datalog-p2p)** *The data complexity of query answering for positive queries in a Datalog-p2p system is in PTIME, while the node complexity of query answering a Datalog-p2p system is EXPTIME-complete.*

*Proof.* The proof is obtained by reducing the problem to a global Datalog program and considering complexity results for Datalog

It can be shown that the node complexity becomes polynomial under the realistic assumption that the number of coordination rules is logarithmic with respect to the number of nodes.

### 3.2 A Distributed Algorithm for Datalog-p2p Systems

Clearly, the global Datalog program devised in the previous Section is not the way how query answering should be implemented in a p2p system. In fact, the global program requires the presence of a *central* node in the network, which knows all the coordination rules and imports all the databases, so that the global program can be executed. A p2p system should implement a *distributed* algorithm, so that each node executes locally a part of it in complete autonomy and it may delegate to neighbour nodes the execution of subtasks, so that there is no need for a centralised authority controlling the process.

In [?] a distributed algorithm for query answering has been introduced, which is sound and complete for an extension of Datalog-p2p systems. In that work, a Datalog-p2p system is called a *definite deductive multiple database*, where domain relations translating query results from the different domains of the various nodes are also allowed. So, we can fully adopt this procedure in our context by assuming identity domain relations. In this paper we do not give the details of the distributed algorithm, which can be found in [?,?].

### 3.3 Acyclic p2p Systems

A p2p system is acyclic if the dependency graph induced by the coordination rules is acyclic. The acyclic case is worth considering since the node complexity of query answering is greatly reduced – it becomes quadratic – and more expressive rules are allowed.

**Theorem 11 (Complexity of acyclic p2p)** *Answering a conjunctive query in an acyclic p2p system with coordination rules having unrestricted conjunctive queries both at the head and at the body is in PTIME. If a positive query is allowed at the head of a coordination rule then query answering becomes coNP-complete. In both cases the node complexity of query answering is quadratic, and it becomes linear in the case of the network being a tree.*

*Proof.* The proof follows by reducing to the problem of query answering using views (see, e.g., [?]). □

This result extends Theorem 3.1 part 2 of [?].

A distributed algorithm for an acyclic p2p system would work as follows. A node answers to a query first by populating the views defined by the heads of the coordination rules of which the node itself is target with the answer to the queries in the body of such rules, and then by answering the query using such views. Of course, answering to the queries in the body of the rules involve recursively the neighbour nodes.

It is possible to exploit the low node complexity of acyclic systems (which have a tree-like topological structure) to build more complex network topologies still with a quadratic node complexity for query answering. The idea is to introduce in an acyclic network the notion of fixed size autonomous subnetworks where cyclic rules are allowed, and a *super-peer* node is in charge to communicate with the rest of the network. This architecture matches exactly the notion of super-peer in real p2p systems like Gnutella.

## 4  Conclusions

In this paper, we propose a new model for the semantics of a p2p database system. In contrast to previous approaches our semantics is not based on the standard first-order semantics.

In our opinion, this approach captures more precisely the intended semantics of p2p systems. It models a framework in which a node can request data from another node, which can involve evaluating a query locally and/or requesting, in turn, data from a third node, but *can not* involve evaluating complex queries over the entire network, as would be the case if the network was an integrated system as in standard work on data integration.

One interesting consequence is in the way we handle inconsistency. In a p2p system, with many independent nodes, there is a possibility that some nodes will contain inconsistent data. In standard approaches, this would result in the whole database being inconsistent, an undesirable situation. In our framework, the inconsistency will not propagate, and the whole database will remain consistent.

The results we have presented show that the original, global, semantics and an alternative, local, semantics are in fact equivalent, and we then extended it in order to handle inconsistency. We also give an algorithm for query evaluation, and some results on special cases where queries can be evaluated more efficiently.

Directions for future work include studying more thoroughly the complexity of query evaluation, as well as special cases, for example ones with appropriate network topologies, for which query evaluation is more tractable. Another issue is that of *domain relations*. These were introduced in [?] to capture the fact that different nodes in a p2p system may not use the same underlying domains, and show how to map one domain to another. Such relations are not studied in the current paper, and their integration in our framework is another area for future research.